\documentclass[floatfix,aps,prd,amsmath,nofootinbib,twocolumn,10pt]{revtex4}

\usepackage{graphicx}
\usepackage{dcolumn}
\usepackage{bm}
\usepackage{epsfig} 
\usepackage{amsfonts}
\usepackage{amsmath}
\usepackage{amssymb}
\usepackage{appendix}
\usepackage[usenames]{color}
\usepackage[dvipsnames]{xcolor}
\usepackage[unicode, colorlinks=true, linkcolor=linkcolor, citecolor=linkcolor, filecolor=linkcolor,urlcolor=linkcolor, pdfusetitle]{hyperref}

\hypersetup{colorlinks,citecolor=blue,linkcolor=blue,urlcolor=blue}
\hypersetup{final=true}

\begin{document}


\title{Non-parametric reconstructions of cosmic curvature: current constraints and forecasts}

\author{Mariana L. S. Dias$^{1}$}
\email{marianadias@on.br}

\author{Ant\^onio F. B. da Cunha$^{1,2}$}
\email{antonioferbapcunha@gmail.com}

\author{Carlos A. P. Bengaly$^{1}$}
\email{carlosbengaly@on.br}

\author{\\ Rodrigo S. Gon\c{c}alves$^{1,2}$}
\email{rsg\_goncalves@ufrrj.br}

\author{Jonathan Morais$^{1,2}$}
\email{jonathanmorais@ufrrj.br}

\affiliation{$^{1}$Observat\'orio Nacional, 20921-400, Rio de Janeiro, RJ, Brazil}

\affiliation{$^{2}$Departamento de F\'isica, Universidade Federal Rural do Rio de Janeiro, 23897-000, Serop\'edica, RJ,  Brazil}

\date{\today}

\begin{abstract}

The assumption of a flat Universe that follows the cosmological principle, i.e., that the universe is statistically homogeneous and isotropic at large scales, comprises one of the core foundations of the standard cosmological model -- namely, the $\Lambda$CDM paradigm. Nevertheless, it has been rarely tested in the literature. In this work, we assess the validity of this hypothesis by reconstructing the cosmic curvature with currently available observations, such as Type Ia Supernova and Cosmic Chronometers. We do so by means of null tests, given by consistency relations within the standard model scenario, using a non-parametric method -- which allows us to circumvent prior assumptions on the underlying cosmology. We find no statistically significant departure from the cosmological principle and null curvature in our analysis. In addition, we show that future cosmological observations, specifically those expected from Hubble parameter measurements from redshift surveys, along with gravitational wave observations as standard sirens, will be able to significantly reduce the uncertainties of current reconstructions.

\end{abstract}

\maketitle

\section{Introduction}
\label{sec:intro}


Since the late 1990s, the standard cosmological model (SCM) corresponds to the flat $\Lambda$CDM scenario. The model states that the Universe is dominated by two components: (i) cold dark matter as the responsible for cosmic structure formation and galaxy dynamics; (ii) the cosmological constant $\Lambda$ as the best candidate to explain the so-called dark energy, i.e., a cosmological component with negative equation of state that is responsible for the cosmic accelerated expansion at late epochs, as initially discovered in~\cite{Riess98, Perlmutter99}. Since then, the $\Lambda$CDM model has shown formidable success to explain cosmological observations, e.g. the Cosmic Microwave Background (CMB)~\cite{Planck:2018vyg}, the luminosity distances of Type Ia Supernovae (SNe)~\cite{Brout:2022vxf,Rubin:2023ovl,DES:2024tys}, the differential galaxy ages by the Cosmic Chronometer (CC) method~\cite{Moresco:2022phi}, in addition to the clustering and lensing of the cosmic large-scale structures (LSS)~\cite{eBOSS:2020yzd,Heymans:2020gsg,DES:2021wwk}. Nonetheless, it suffers from theoretical issues, such as fine-tuning and coincidence problems, as well as observational caveats that appeared in the last decade. The most prominent problem of the latter case is the $\sim5\sigma$ tension between Hubble Constant ($H_0$) measurements obtained from early-time probes, like the CMB, and late-time observations of SNe through the cosmic distance ladder approach. This could be due to unaccounted for systematics and possible biases in at least one of those measurements~\cite{Perivolaropoulos:2024yxv}, although a potential departure of the SCM cannot be ruled out -- see~\cite{DiValentino:2021izs,Hu:2023jqc} for a broader discussion on this topic, and~\cite{Perivolaropoulos:2021jda} for other possible challenges of the $\Lambda$CDM. 

Given such a scenario, it is extremely important to put the physical foundations of the standard model under scrutiny, as any statistically significant deviation from them would require a complete reformulation of our cosmological understanding. These foundations consist of the theory of general relativity as the underlying theory of gravity, and the assumption of statistical large-scale homogeneity and isotropy at large scales. The latter corresponds to the cosmological principle (CP), which allows us to describe cosmic distances and clocks by the Friedmann-Lema\^itre-Robertson-Walker (FLRW) metric. Although the former has been extensively probed in local laboratories, as well as in cosmic scales, tests of the latter assumption are much more scarce -- we refer the reader to~\cite{Aluri:2022hzs} for a review on these tests. Moreover, we remark that only the assumption of cosmic isotropy can be directly tested with cosmological observations, conversely from the homogeneity assumption, because we can only carry out such observations along the past light-cone, rather than time-constant hypersurfaces~\cite{Clarkson:2010uz,Maartens:2011yx,Clarkson:2012bg}.

Still, some consistency tests have been designed to check the validity of the CP in such a way that can circumvent this limitation on testing cosmic homogeneity. One of them corresponds to a consistency relation between cosmic distances and clocks, along with their respective derivatives, that must be respected in a FLRW Universe, as originally proposed in~\cite{Clarkson:2007pz}. In practice, one verifies whether there is any evolution of the cosmic curvature parameter across different redshift ranges -- if that is indeed the case, then the CP will be immediately ruled out. Similar tests have been developed and carried out in~\cite{Clarkson:2007bc,Sapone:2014nna,Rasanen:2014mca,Cai:2015pia,Cai:2016vmn,Wang:2017lri,Yu:2016gmd,Collett:2019hrr,Yang:2020bpv,Mukherjee:2022ujw,Wu:2022fmr}, where none of them reported significant deviations for the CP assumption. These results are complemented by other analyses focused on measuring the cosmic homogeneity scale in the large-scale distribution of galaxies or quasars in the Universe, as performed in~\cite{Laurent:2016eqo, Ntelis:2017nrj, Goncalves:2017dzs, Goncalves:2018sxa, Goncalves:2020erb, Andrade:2022imy, Goyal:2024ctd}, where the mentioned scale has been positively identified. 

Nevertheless, it is important to revisit such a test in light of the most up-to-date cosmological observations from SNe and CC, as they consist of standardisable candles and chronometers -- quantities that are needed for this purpose. So, our goal in this work is to obtain updated constraints on the FLRW null test, i.e., we test the behaviour of cosmic curvature versus redshift, as in~\cite{Clarkson:2007pz}, as well as a variation of the same test presented in~\cite{Cai:2015pia}, which provides a null test of the flat curvature. Also, we forecast the precision that can be achieved with a combination of upcoming observational data that has not been explored yet in previous works, as presented in~\cite{Cai:2015pia,Wu:2022fmr}, and that can be realistically achieved within the next decade. More specifically, we simulate Hubble parameter measurements with the precision that is expected by redshift surveys like Javalambre Physics of the Accelerating Universe Astrophysical Survey (J-PAS)~\cite{J-PAS1}, in terms of measuring the radial mode of baryon acoustic oscillations (BAO), along with gravitational waves as standard siren measurements that are expected by interferometers such as LIGO~\cite{LIGO}. As a final remark, we note that our analysis is performed using a non-parametric method, in order to make minimal assumptions on the underlying cosmological model.

This paper is organised as follows: Section 2 describes the theoretical framework and the tests to be carried out in our work; Section 3 presents the observational and simulated datasets; Section 4 describes the method used to perform non-parametric reconstructions; Section 5 shows the results obtained in our analysis; Section 6 contains the concluding remarks. 

\section{Cosmological Tests for Cosmic Curvature}
\label{sec:CosmoTest}

\subsection{Basic quantities}

The most general metric describing an expanding homogeneous and isotropic universe is the FLRW metric which is written in spherical coordinates as
\begin{equation}
ds^2 = c^2dt^2 - a^2(t) \left( \frac{dr^2}{1 - k r^2} + r^2d\theta^2 + r^2 \sin^2{\theta} d\phi^2 \right),
\end{equation}
where $c$ is the speed of light, $a(t)$ is the scale factor and $k$ is a constant that defines the spatial curvature of the Universe and can only assume the values $k = -1, 0, +1$.

When we assume that the Universe is composed by a perfect-like fluid, the combination of the FLRW metric and the Einstein Field's Equation of general relativity lead to the Friedmann's equations, where the previously mentioned constant is related with the cosmic curvature in a certain redshift as
\begin{equation}
\Omega_k (z) \equiv - \frac{k c^2}{a^2(z) H^2(z)},
\end{equation}
where $H(z)$ is the Hubble parameter and for the present value of it (the so called Hubble parameter, $H_0$) the previous equation becomes
\begin{equation}
\Omega_{k,0} = - \frac{k c^2}{H_0^2},
\end{equation}
and the three possible values for the current cosmic curvature parameter (and the respective characterisations) are
\begin{equation}
\Omega_{k,0}
\begin{cases}
> 0 \; (open) \\
= 0 \; (flat) \\
< 0 \; (closed) 
\end{cases}
\end{equation}
Moreover, we would like to note that, although the currently available observations of CMB, SNe and LSS are consistent with a flat case ($\Omega_{k,0}=0$) within a $10^{-3}$ precision, some of these probes cannot individually constrain this parameter with such a high precision -- and some of them show results which are inconsistent with a flat Universe, as in the case of SNe. So, one should be cautious when setting $\Omega_{k,0}=0$ in a parametric analysis, as discussed in~\cite{Anselmi:2022uvj}. 

The cosmological parameters can be inferred from the observation of cosmological distances as, for instance, the luminosity distance($d_L$). With effect, the expression for $d_L$ obtained from the FLRW metric is
\begin{equation}
\label{LumDist}
d_L(z) = \frac{c(1+z)}{H_0 \sqrt{ \vert \Omega_{k,0} \vert}} \mathcal{F} \left( \sqrt{\vert \Omega_{k,0} \vert} \int_0^z \frac{dz'}{E(z')} \right),
\end{equation}
where $E(z) \equiv H(z)/H_0$ is the reduced Hubble parameter and $\mathcal{F}$ is a function that assumes different functional forms depending on the value of $\Omega_{k,0}$ as
\begin{equation}
\mathcal{F}(x) =
\begin{cases}
sinh(x), \; \; \mathrm{if} \; \; \Omega_{k,0} > 0 \\
x, \; \; \mathrm{if} \; \; \Omega_{k,0} = 0 \\
sin(x), \; \; \mathrm{if} \; \; \Omega_{k,0} < 0
\end{cases}
\end{equation}

We can also define the dimensionless comoving distance ($D(z)$) from the luminosity distance as
\begin{equation}
\label{DimlssDist}
D(z) = \frac{H_0 d_L(z)}{c (1+z)}.
\end{equation}

In order to test whether the cosmic curvature of the Universe is currently zero as well as zero at any redshift, we perform the following two null tests.

\subsection{Null tests for cosmic curvature}

Following~\cite{Clarkson:2007pz,Clarkson:2007bc}, from Eq.~\ref{LumDist} we have
\begin{equation}
\label{OmegaKTest}
\Omega_{k,0} = \frac{E^2(z) D'^2(z) - 1}{D^2(z)},
\end{equation}
where $D'$ is the redshift derivative of Eq.~\ref{DimlssDist}. So we have the first null condition as: $\Omega_{k,0} \neq \mbox{constant} \Rightarrow$ implies FLRW ruled out.

Besides, we can rewrite Eq.~\ref{OmegaKTest} and obtain
\begin{equation}
\frac{\Omega_{k,0} D^2(z)}{E(z) D'(z) + 1} = E(z) D'(z) - 1,
\end{equation}
then, following~\cite{Cai:2015pia} we define
\begin{equation}
\label{OkTest1}
\mathcal{O}_k (z) \equiv E(z) D'(z) - 1.
\end{equation}
As this quantity is always zero at $z = 0$, since $E(z=0) = 1$ and $D'(z) = 1/E(z)$, then we have the second null condition as: $\mathcal{O}_k(z) \neq 0$ implies that a flat universe is ruled out at any non-zero redshifts~\cite{Cai:2015pia}. It is worth noticing that the first null condition is valid for $z = 0$, while the second null condition regards to $z > 0$. We also stress that the $E(z)$ and $D(z)$ measurements we hereafter use in our analysis are independent of each other, as they come from different observables, as we detail in the next section.

\section{Data}
\label{sec:obs_data}

\subsection{Current observations}

We adopt one of the latest SN compilations, namely the Pantheon+ and SH0ES data-set~\cite{Brout:2022vxf} (see also~\cite{Scolnic:2021amr,Riess:2021jrx}), which provides 1701 light curve measurements of 1550 distinct SNe in the redshift interval $0.001 < z < 2.26$. Hence, we have 1701 measurements of SN apparent magnitudes as a function of redshift, $m_{B}(z)$, which can be combined with the determination of the SN absolute magnitude given by
\begin{equation}\label{eq:MB}
    M_{B} = -19.25 \pm 0.03 \,,
\end{equation}
so that we can obtain the luminosity distances according to
\begin{equation}\label{eq:DLz}
    d_{L}(z)=10^{\frac{m_{B}(z)-M_{B}-25}{5}} \,,
\end{equation}
which relates to $D(z)$ through Eq.~\ref{DimlssDist}. 

On the other hand, our $H(z)$ data consist of a compilation of 31 measurements, as displayed in~\cite{Bengaly:2022cgs} -- see also~\cite{Wang:2019vxv} and references therein -- obtained from differential ages of passively evolving galaxies via the relation $H(z) = -(1+z)^{-1} dz/dt$, i.e., the so-called the cosmic chronometer method~\cite{Jimenez:2001gg}. For further details about the cosmic chronometer method, we refer the interested reader to~\cite{Moresco:2024wmr}.

Note that we chose not to include the $H(z)$ data points from radial BAO mode measurements, as those shown in Table 2 of~\cite{Wu:2022fmr}, because they implicitly assume the $\Lambda$CDM model as the fiducial cosmological model -- which could bias our results in favor of the standard model. However, this is not an issue for simulated datasets since they require the assumption of a fiducial model anyway, as we detail now.

\subsection{Simulations of Hubble Parameter from Redshift Surveys}

We also simulate future $H(z)$ measurements obtained from the radial mode of baryon acoustic oscillations \cite{Gaztanaga:2008xz} that are expected to be observed by ongoing redshift surveys, as the case of the Javalambre Physics of the Accelerating Universe Astrophysical Survey, namely J-PAS. We produce 23 data points which roughly follows the number of $H(z)$ measurements provided by Sloan Digital Sky Survey (see table 2 in~\cite{Wu:2022fmr}) from a realisation of the flat $\Lambda$CDM model according to $H(z) \rightarrow \mathcal{N}(H^{\rm fid}(z), \sigma_{H(z)})$, so that the Hubble parameter follows the Friedmann equation
\begin{equation}\label{eq:hz_fid}
E^{\rm fid}(z) \equiv \left[\frac{H^{\rm fid}(z)}{H^{\rm fid}_0}\right]^2 = \Omega^{\rm fid}_{\rm m}(1+z)^3 + (1-\Omega^{\rm fid}_{\rm m}) \,.
\end{equation} 
Here, we assume $\Omega^{\rm fid}_{\rm m} = 0.334$ and $H^{\rm fid}_0 = 73.6 \, \mathrm{km \, s}^{-1} \, \mathrm{Mpc}^{-1}$ as our fiducial cosmology, which agrees with the flat $\Lambda$CDM best-fit obtained from the Pantheon+ and SH0ES SN compilation~\cite{Brout:2022vxf}, and that the $H(z)$ uncertainties, $\sigma_{H(z)}$, should follow the values shown in Fig. 15 of~\cite{AparicioResco:2019hgh} for a J-PAS 8500 degree$^2$ configuration. Also, we assume that these data points should follow a redshift distribution $P(z)$ across the redshift range $0.3<z<2.5$ such as~\cite{Wang:2019vxv,Bengaly:2022cgs}
\begin{equation}\label{eq:pz}
P(z; \; k,\theta) = z^{k-1}\frac{e^{-z/\theta}}{\theta^{k}\Gamma(k)} \;,
\end{equation}
where we fix $\theta$ and $k$ to their respective best fits to the real data, i.e., $\theta_{\rm bf}=0.647$ and $k=1.048$~\cite{Bengaly:2022cgs}.   

\subsection{Simulations of Luminosity Distance from Gravitational Waves}


By assuming the same fiducial model, we also performed simulations of luminosity distances from gravitational wave events (GW). Due to the redshift interval we specify the sources of the events as mergers of Neutron Stars-Black Hole (NS-BH) or Neutron Stars-Neutron Stars (NS-NS) binaries. They obey a redshift distribution given by
\begin{equation}\label{eq_pz}
    P(z) \propto \frac{4 \pi H_0 d_{c}^{2}R(z)}{E(z)(1+z)},
\end{equation}
where $d_c \equiv (c/H_0) D$ is the comoving distance, the factor $R(z)$ describes the evolution of the star formation rate and it has a specific functional form for each gravitational wave source \cite{Zhao:2011kcl}. 

From the probability distribution for each gravitational source we randomly pick 1000 points in the redshift range $0 < z < 2.5$ and, due to the redshift interval, the type of gravitational wave sources are bright sirens \cite{Abbott:2022grw}. The choice of  $\mathcal{O}(10^3)$ events is motivated by the expected number of detections in the next ten years \cite{Cai:2016sby}. We then obtain the luminosity distance at those redshift with the fiducial model, and we perform a Monte Carlo simulation by assuming a gaussian distribution centered on these fiducial $d_L$. The standard deviation is defined upon the instrumental error, that is related with the signal-to-noise ratio ($SNR$) of the detection, the amplitude of the gravitational wave, the interferometer antenna pattern ($F$) and the power spectrum density ($PSD$). The uncertainty is combined with an additional error from weak lensing (for a detailed description about this procedure see \cite{Zhang:2019dfh} and references therein). We simulate our data with a $F$ and $PSD$ based on a specific interferometer pattern as the LIGO collaboration~\cite{Schutz:2011dgh}.

Furthermore, we remark that we choose these GW measurements as a probe of these null tests because we expect a larger number of data points at higher redshifts (e.g. $z>1$) comparing to SNe -- see figures in the Appendix \ref{sec:app} for a better understanding of redshift distribution difference between GW and SNe.

\section{Gaussian Process method}
\label{sec:GP}

The quantities we need to reconstruct from the observed and simulated datasets are $E(z)$, $D(z)$, and $D'(z)$, so that we can compute $\Omega_{k,0}$ and $\mathcal{O}_k (z)$ as in Eqs.~\ref{OmegaKTest} and~\ref{OkTest1}, respectively, as well as their correspondent uncertainties. Because we want to circumvent {\it a priori} assumptions about the cosmological model that describes the relation between the Hubble parameter and luminosity distance with the redshift, we deploy a non-parametric approach using the Gaussian Process (GP) method.

By definition, a GP consists of a distribution over functions that can best describe the patterns of the available data. Hence, we can compute any quantity of interest without assuming an explicit parametrisation. Here, we use the well-known {\sc GaPP} (Gaussian Processes in Python) package~\cite{Seikel:2012uu}\footnote{\url{https://github.com/astrobengaly/GaPP}} for this purpose.

We compute the uncertainties of $E(z)$, $D(z)$, and $D'(z)$ from the values provided by the {\sc GaPP} code after optimising the GP hyperparameters, assuming the squared exponential kernel (unless stated otherwise) for 1000 uniformly spaced-out bins at the $0<z<2.5$ range, so that we obtain the uncertainties of $\Omega_{k,0}$ and $\mathcal{O}_k (z)$ by error-propagating them with respect to $E(z)$, $D(z)$, and $D'(z)$. 

\section{Results}
\label{sec:results}

\begin{figure}[!t]
    \centering
    \includegraphics[width=0.9\linewidth]{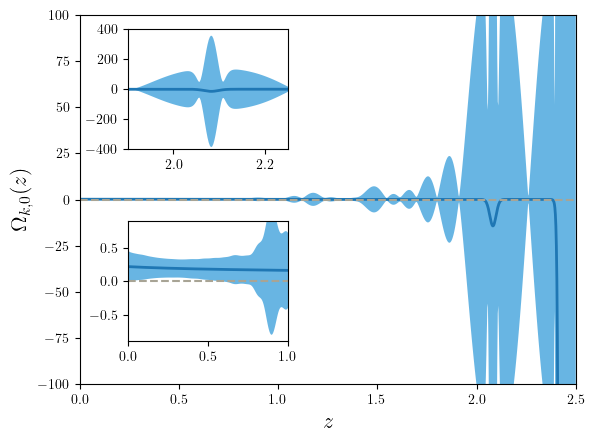}  
    \includegraphics[width=0.9\linewidth]{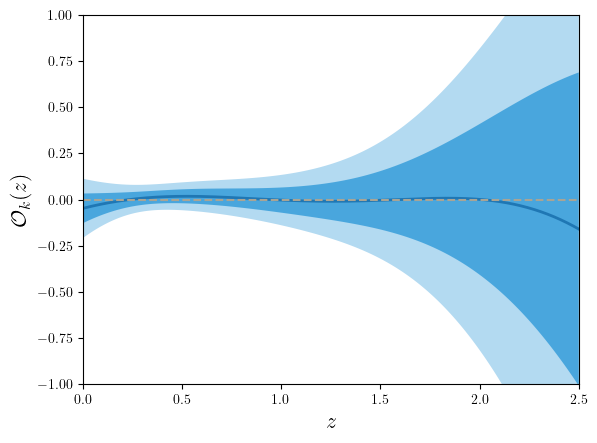}  
    \caption{In the top panel, we show the results for $\Omega_{k,0}$ as a function of redshift, based on reconstructions from observational data, with its respective uncertainty at $1\sigma$ confidence level. In the upper inset, there is a zoomed-in view of the redshift range $1.9<z<2.25$, while the lower inset shows a zoomed-in view of the redshift range $0<z<1$, along with the dashed gray line representing $\Omega_{k,0} = 0$ (flat Universe). Both highlighted zoomed-in views emphasize the behaviour of uncertainties within their respective ranges. In the bottom panel, we have the evolution of $\mathcal{O}_k(z)$ as a function of redshift, also based on reconstructions from observational data, with its respective uncertainty at $1\sigma$ and $2\sigma$ confidence levels. The solid navy line stands for the mean of the reconstruction and the dashed gray line stands for the case $\mathcal{O}_k(z) = 0$.}
    \label{fig:omegak_sneia_cc}
\end{figure}

\begin{figure}[!t]
    \centering
    \includegraphics[width=0.9\linewidth]{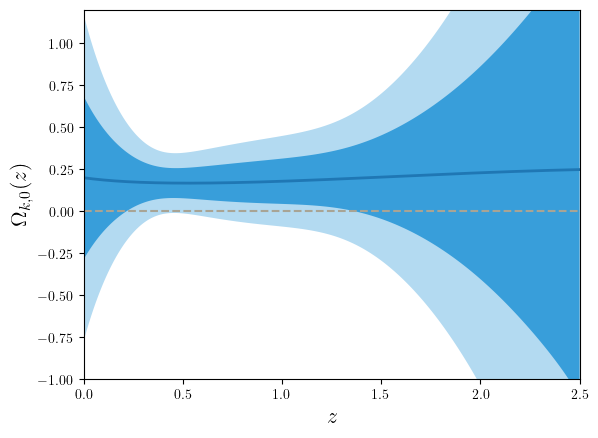}
    \includegraphics[width=0.9\linewidth]{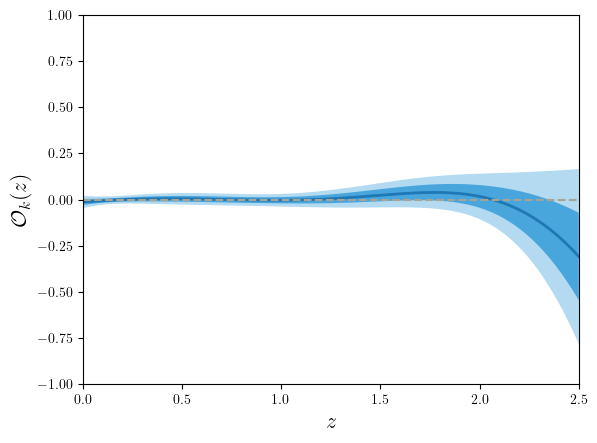}
    \caption{In the top panel, we show the evolution of $\Omega_{k,0}$ as a function of redshift, for $1$ and $2\sigma$ confidence levels, based on reconstructions from simulated data. In the bottom panel, we show the evolution of $\mathcal{O}_k(z)$ as a function of redshift, also for $1$ and $2\sigma$ confidence levels, based on reconstructions from simulated data. In both panels, the dashed gray line represents the flat Universe case, i.e., $\Omega_{k,0}=0$ and $\mathcal{O}_k(z)=0$.}
    \label{fig:omegak_ligo_jpas}
\end{figure}

\begin{figure}[!t]
    \centering
    \includegraphics[width=0.9\linewidth]{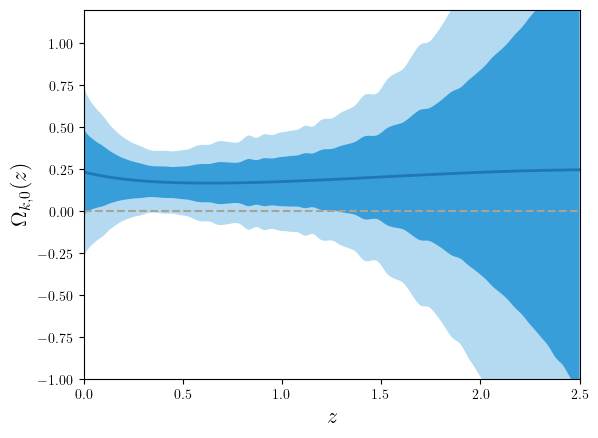}
    \includegraphics[width=0.9\linewidth]{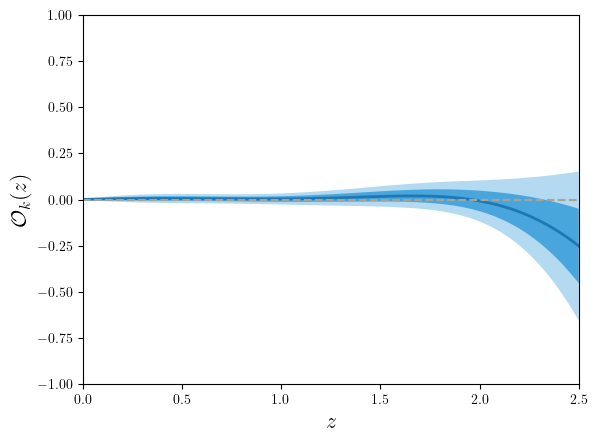} 
    \caption{Same as Fig.~\ref{fig:omegak_ligo_jpas}, but rather for the joint analysis using current and future observations.}
    \label{fig:omegak_combined}
\end{figure}

We perform the null tests mentioned in Eqs.~\ref{OmegaKTest} and \ref{OkTest1} using the GP method, and the previously mentioned datasets, so that the results obtained for those null tests are described as follow

\subsection{Observational data}

In the top panel of Fig.~\ref{fig:omegak_sneia_cc}, we show the results of $\Omega_{k,0}$ with $1\sigma$ confidence level -- we refer the reader to the Appendix \ref{sec:app} for the numerical reconstructions of $D(z)$ and $E(z)$, see Fig.~\ref{fig:GPRecObs}. As we can see, the value of the cosmic curvature parameter is compatible with a flat universe, i.e., $\Omega_{k,0}=0$. The uncertainties increase significantly at redshifts $z>1.5$ (see the top inset in this Figure) compared to lower redshift ranges, such as $0<z<1$ (bottom inset in the same Figure). This is expected because of the scarcity of observational data in that redshift range, which degrades the reconstruction. Furthermore, we note a constancy in the values of $\Omega_{k,0}$ across the $0<z<2.5$ interval, which thus indicates that there is no statistically significant deviation from the assumption of the FLRW universe, i.e. the first null condition.

In the bottom panel of Fig.~\ref{fig:omegak_sneia_cc}, we show the results obtained for the $\mathcal{O}_k(z)$ null test, along with their correspondent $1\sigma$ and $2\sigma$ confidence level uncertainties. The mean curve (navy solid line) and the null condition $\mathcal{O}_k(z) = 0$ (gray dashed line) are also shown in the figure. These results are again compatible with a flat universe, hence indicating no hints at a possible departure from the second null condition. 

In addition, we note that we assume $H_0$ values on the $D(z)$ and $E(z)$ reconstructions that are directly obtained from the SN and CC data, respectively, as they consist of independent datasets. Namely, we assume $H_0 = 73.6 \pm 1.1 \, \mathrm{km \, s}^{-1} \, \mathrm{Mpc}^{-1}$ for the former, as measured by SH0ES, and $H_0 = 67.4 \pm 4.7 \, \mathrm{km \, s}^{-1} \, \mathrm{Mpc}^{-1}$ for the latter, as obtained through the GP extrapolation of the $H(z)$ data points at $z=0$. We also propagate the $H_0$ uncertainties through standard error deviation in both cases.

\subsection{Simulations}

In order to quantify the impact of the next generation instruments in those null tests, we also perform other analyses with the simulations described in the previous section, that being Hubble parameter measurements by ongoing redshift surveys, like J-PAS, and GW standard siren measurements from LIGO. Note that, differently from the previous case, we assume the same $H_0$ value for both J-PAS $E(z)$ and LIGO $D(z)$ reconstructions, that being $H_0 = 73.6 \pm 1.1 \, \mathrm{km \, s}^{-1} \, \mathrm{Mpc}^{-1}$, as we assumed the SH0ES measurement as the fiducial cosmological model to produce both simulated datasets.

The results obtained for both null tests can be found in Fig. \ref{fig:omegak_ligo_jpas}, whereas the numerical reconstructions of $D(z)$ and $E(z)$ are also displayed in the Appendix \ref{sec:app}, see Fig.~\ref{fig:GPRecSim}. We find that the relative uncertainties of the $\Omega_{k,0}$ (top panel) and $\mathcal{O}_k(z)$ (bottom panel) reconstructions are significantly reduced compared to the current results, especially at higher redshifts, $z>1.0$, where they decrease by over an order of magnitude. This occurs due to the larger number of $D(z)$ data points from GW at intermediate to higher redshifts (around $z>0.5$) compared to the SNe, which reflects in much more restricted uncertainties in this case. 


For the sake of completeness, we also performed a joint analysis with both observational and simulated data (Fig. \ref{fig:omegak_combined}). The colors and confidence intervals follow the pattern of the previous plots. As we can see, both null tests give similar results to those displayed in Fig.~\ref{fig:omegak_ligo_jpas}, albeit with reduced uncertainties at the low redshift end, due to the presence of SN data points.

\section{Conclusions}
\label{sec:conclusions}

In this work, we search for possible evidences of an evolution in the cosmic curvature as a function of redshift. We deploy two parameters for this purpose: (i) The $\Omega_{k,0}$ parameter, which tests the current value of the curvature parameter; (ii) The $\mathcal{O}_k$ parameter, which analyses whether the curvature is zero at any other redshift. If we detect a redshift evolution was in the first parameter, it would imply that the FLRW metric is not able to describe the observable Universe, thus being the null hypothesis behind this test. As for the second parameter, such an evolution would indicate that the observable universe is not flat, which is its null hypothesis. Hence, if we fail to reject any of these null hypotheses are rejected, we would have an immediate evidence of new physics beyond the standard cosmological model -- especially given that the first null test consist on one of the most fundamental pillars of the standard model, i.e., the cosmological principle. 

Hence, by means of a non-pametric, Gaussian Process-based analysis, on currently available data from Type Ia supernovae (Pantheon+ with SH0ES) and differential galaxy ages (cosmic chronometers, CC), we found no evidence for a non-flat Universe, as well as no evidence supporting any evolution of the current curvature parameter, thereby rejecting the null hypothesis of a non-FLRW or a non-flat universe. Nonetheless, the uncertainties in our observational analysis are large, specially at higher redshifts ($z>1)$ due to the scarcity of the data at those ranges. So, we forecast how much these uncertainties can be improved in light of cosmological data that are plausible to be obtained within the next decade. For this purpose, we produced simulated data from gravitational waves and baryon acoustic oscillations following LIGO and J-PAS specifications, respectively, where we found out that this combination can indeed yield much improved uncertainties in both null tests. Notably, we could reach an improvement of over an order of magnitude at $z>1.0$ in both cases, which is mostly due to the larger amount of expected gravitational wave standard siren measurements located at this redshift range.

This result demonstrates that this specific combination of redshift surveys with gravitational waves has a great potential to improve the precision of crucial consistency tests of the standard model, as those related to the cosmic curvature. A more thorough analysis of what could be the optimal combination of future cosmological experiments on those tests will be explored in a future work. Still, our results confirm that there is no evidence of deviation from one of the foundations of the standard cosmological model considering the limitations of current observational data, which helps underpinning its validity. 



{\it Acknowledgments:} 
CB acknowledges financial support from Funda\c{c}\~ao \`a Pesquisa do Estado do Rio de Janeiro (FAPERJ) - Postdoc Recente Nota 10 (PDR10) fellowship. AFBC thanks financial support from Universidade Federal Rural do Rio de Janeiro (UFRRJ) grant PICE4154-2023 (PIBIC/UFRRJ) and Observatório Nacional (ON) grant 1683712 (PICT/ON). RSG thanks financial support from the Fundação de Amparo à Pesquisa do Estado do Rio de Janeiro (FAPERJ) grant SEI-260003/005977/2024 - APQ1. MLSD thanks CAPES for financial support. JM acknowledges financial support from Conselho Nacional de Desenvolvimento Científico e Tecnológico (CNPQ) - Undergraduate research fellowship.

\newpage



\onecolumngrid
\appendix{Appendix}


\newpage


\section{Reconstructions of $E(z)$, $D(z)$ and $D'(z)$}\label{sec:app}
\label{sec:GPfunctions}

For the sake of completeness, we present below the reconstruction curves obtained for the functions $E(z)$, $D(z)$ and $D'(z)$ for the observational and simulated data.

\begin{figure*}[!h]
	\centering
	\includegraphics[width=0.49\textwidth]{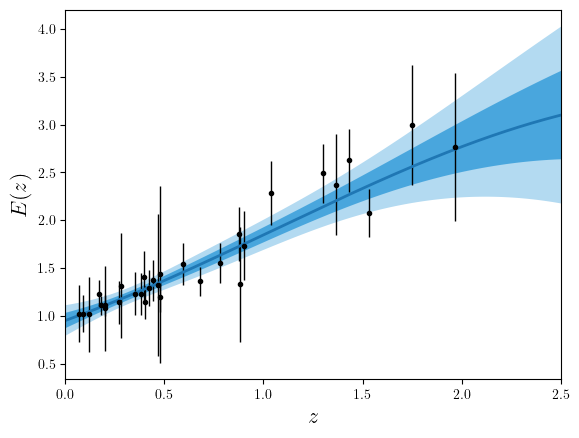}
        \includegraphics[width=0.49\textwidth]{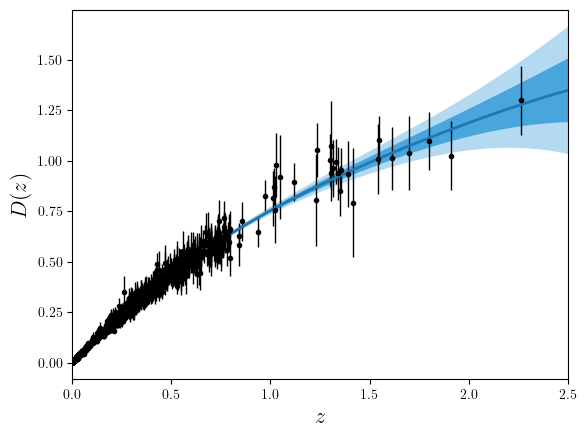}
        \includegraphics[width=0.49\textwidth]{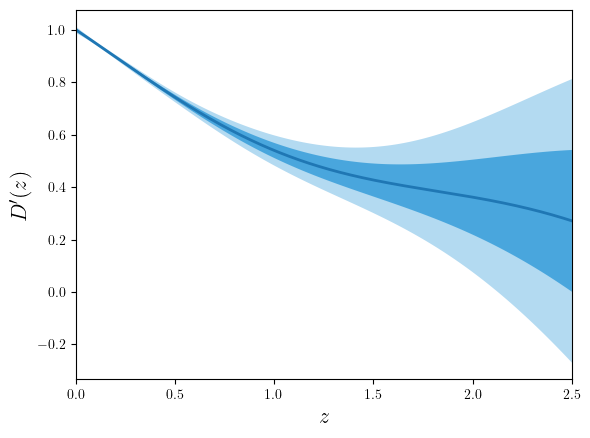}
	\caption{Reconstructions of $E(z)$ (cosmic chronometers), $D(z)$ (SNe) and $D'(z)$ functions for the observational data.}
	\label{fig:GPRecObs}
\end{figure*}

\begin{figure*}[!h]
	\centering
	\includegraphics[width=0.49\textwidth]{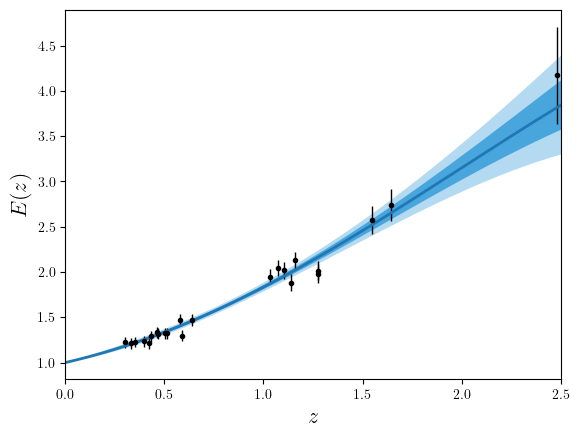}
        \includegraphics[width=0.49\textwidth]{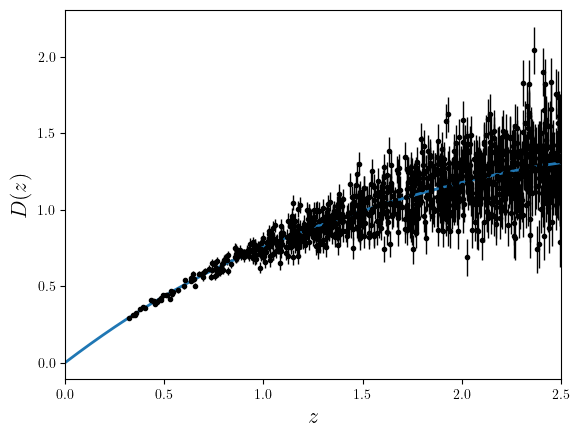}
        \includegraphics[width=0.49\textwidth]{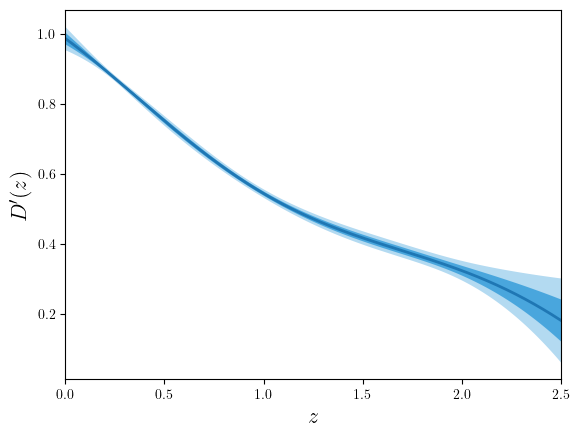}
	\caption{Reconstructions of $E(z)$ (redshift surveys), $D(z)$ (GW) and $D'(z)$ functions for the simulated data.}
	\label{fig:GPRecSim}
\end{figure*}

\end{document}